\documentclass[]{elsarticle}
\usepackage{lineno,hyperref}
\usepackage{epstopdf} 
\usepackage{color}
\usepackage{lscape}
\modulolinenumbers[5]
\biboptions{sort&compress}
\journal{}
\usepackage{graphicx}
\usepackage{amssymb}
\begin{document}
\begin{frontmatter}
\title{Relation of the thermodynamic parameter of disordering with the width of structure factor and defect concentration in a metallic glass}

\author[VSPU]{G.V. Afonin}
\author[VSPU]{A.S. Makarov}
\author[VSPU]{R.A. Konchakov}
\author[NWPU]{J.C. Qiao}
\author[MISiS]{A.N. Vasiliev}
\author[IFTT]{N.P. Kobelev}
\author[VSPU]{V.A. Khonik}
\corref{cor}
\cortext[cor]{Corresponding author} 
\ead{v.a.khonik@yandex.ru} 
\address[VSPU] {Department of General Physics, Voronezh State Pedagogical
University,  Lenin St. 86, Voronezh 394043, Russia}
\address[NWPU] {School of Mechanics, Civil Engineering and Architecture, Northwestern Polytechnical University, Xi’an 710072, China}
\address[MISiS] {National University of Science and Technology MISiS, Moscow 119049, Russia}
\address[IFTT] {Institute of Solid State Physics RAS, Chernogolovka, 142432, Russia}

\begin{abstract}

In this work, we show that above the glass transition there exists a strong
unique interrelationship between the thermodynamic parameter of disorder of
a metallic glass derived using its excess entropy, diffraction measure of disorder
given by the width of the X-ray structure factor and defect concentration derived
from shear modulus measurements. Below the glass transition, this relationship
is more complicated and depends on both temperature and thermal prehistory.
\end{abstract}


\end{frontmatter}

Understanding of general key features of relaxation behavior of metallic glasses (MGs) is of vital importance for the adequate description of their structural state and atomic rearrangements \cite{ChengProgMatSci2011,WangProgMaterSci2019}. In particular, current literature gives quite a few examples for the interpretation of relaxation properties in terms of different types of defects assumed to exist in real MGs \cite{Ye,Egami,Betancourt,LiSciRep2015,Zhang,WangNatSciRev2018,WangProgMaterSci2019}. However, any relation of the defect structure with macroscopic structure parameters remains mostly out of view. An attempt to advance in this direction was recently reported in our work \cite{MakarovInmermetallics2023}, in which it was found that the width of the structure factor given by the full width at half maximum (FWHM) of the first peak of structure factor is related to  the concentration $c_{def}$ of interstitial-type defects  derived from shear modulus measurements of a Pt-based metallic glass. It was shown, in particular,   that above the glass transition temperature $T_g$, the FWHM linearly increases with $c_{def}$. This indicates that heat absorption and related increase of $c_{def}$ provide a significant defect-induced disruption of the dominant short-range order increasing thus the integral structural disordering. Below $T_g$, the relationship between the FWHM and $c_{def}$ is more complicated but defect-induced ordering is observed upon approaching $T_g$.  In any case, it is clear that the diffraction measure of disorder is related to the defect concentration.  

On the other hand, within the view of statistical physics \cite{Landau}, any ordering/disordering should be related to a change of glass entropy. This approach was developed in our recent work \cite{MakarovScrMater2023}, in which a dimensionless thermodymanic parameter of structural order was introduced as $\xi(T)=1-\Delta S(T)/\Delta S_{melt}$, where $\Delta S(T)$ is temperature-dependent excess entropy of glass with respect to the maternal crystal and $\Delta S_{melt}$ is the entropy of melting. The above order parameter  varies in the range $0<\xi<1$ upon changing the structure from the fully disordered (liquid-like) state with $\Delta S \rightarrow \Delta S_{melt}$ and $\xi \rightarrow 0$ towards the fully ordered state (crystal-like) characterized by $\Delta S \rightarrow 0$ and $\xi\rightarrow 1$. An application of this approach to 13 metallic and 2 tellurite glasses showed that the order parameter $\xi$ provides very useful qualitative and quantitative information  on structural ordering of various glasses and its evolution upon  relaxation. This order parameter is quite sensitive to the glass state and its chemical composition. In particular, structural relaxation below $T_g$ results in a significant increase of $\xi$- parameter while heating above $T_g$ leads to its rapid decrease. It was also found that 
 order parameter $\xi_{sql}$ calculated for the supercooled liquid state is related to the glass forming ability (GFA). An increase of $\xi_{sql}$ strongly worsens the GFA. 

While the parameter $\xi$  describes the order in glass, the quantity 
\begin{equation}
\alpha(T) =\frac{\Delta S(T)}{\Delta S_{melt}}\label{alpha}
\end{equation}
describes structural disordering. In this work, we determined the parameter $\alpha$ for the aforementioned Pt-based glass and found that it is strongly related to the FWHM and $c_{def}$ indicating thus that the thermodynamic entropy-based and diffraction characteristics of disordering are \textit{i}) interrelated and \textit{ii}) linked to the defect concentration.

We studied the same glass Pt$_{42.5}$Cu$_{27}$Ni$_{9.5}$P$_{21}$ (at.\%) produced in the bulk form  by melt jet quenching \cite{MakarovInmermetallics2023} while its FWHM and defect concentration $c_{def}$ as  functions of temperature were also taken from this work. The excess entropy of glass with respect to the maternal crystal was calculated from calorimetric measurements as \cite{MakarovScrMater2023,MakarovJPCM2021}
\begin{equation}
\Delta S(T)=\frac{1}{\dot{T}}\int_{T}^{T_{cr}} \frac{\Delta W(T)}{T}dT, \label{DeltaS}
\end{equation}   
where $\Delta W(T)=W_{gl}(T)-W_{cr}(T)$ is the differential heat flow, $W_{gl}(T)$ and $W_{cr}(T)$ are the heat flows coming from glass and its maternal crystal, respectively, $\dot{T}$ is the heating rate and $T_{cr}$ is the temperature of the complete crystallization. Specific details of calorimetric measurements are given in Ref.\cite{MakarovScrMater2023}. These measurements were performed at a heating rate of 3 K/min on both initial and relaxed samples, where those latter were prepared by heating to 559 K (deep in the supercooled liquid state) and cooling back to room temperature at the same rate. Note that if  current temperature $T=T_{cr}$ then $\Delta S=0$ and, therefore, the function $\Delta S (T)$  represents temperature dependence the excess entropy of glass with respect to the maternal crystal. The melting entropy  $\Delta S_{melt}$ was accepted to be  11.3 J/(K$\times$mol) \cite{NeuberActaMater2021}. This allows calculating the parameter of disordering $\alpha$, which changes from $\alpha \rightarrow 1$ for the fully disordered (liquid-like) state to $\alpha \rightarrow 0$ for fully ordered (crystal-like) state. 

\begin{figure}[t]
\center{\includegraphics[scale=0.35]{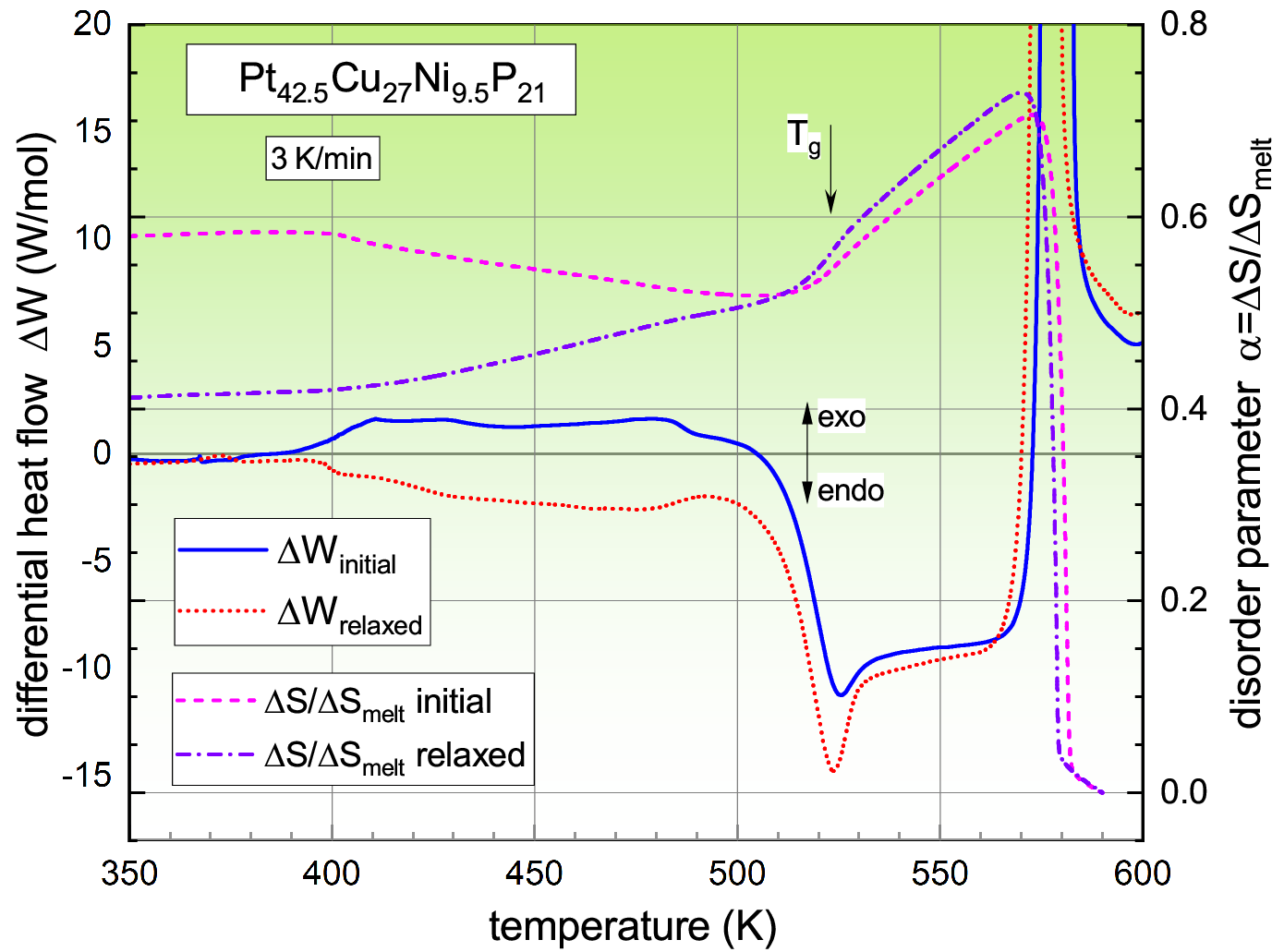}}
\caption[*]{\label{Fig1.eps} Differential heat flow $\Delta W$ and disorder parameter $\alpha=\Delta S/\Delta S_{melt}$ for the glass under investigation in the initial and relaxed states, where the excess entropy of glass with respect to maternal crystal $\Delta S$ is calculated using Eq.(\ref{DeltaS}).}
\end{figure} 

Figure \ref{Fig1.eps} gives temperature dependences of the differential heat flow $\Delta W$ for the glass under investigation in the initial and relaxed states. These curves were  used to calculate the excess entropy $\Delta S$ with Eq.(\ref{DeltaS}) and next determine 
the disorder parameter $\alpha=\Delta S/\Delta S_{melt}$, which is also given in Fig.\ref{Fig1.eps}. It is seen that exothermal heat flow of the initial sample in the range 400 K$<K<$510 K (below $T_g$) leads to a decrease of disorder parameter $\alpha$ in the same range, just as one would expect. Continued heating leads to a strong heat absorption near and above $T_g$ (i.e. in the supercooled liquid state), which is accompanied by a marked increase of disorder  (as anticipated) determined by $\alpha$-increase from 0.52 to 0.71 in the same range. Relaxed sample instead of exothermal flow displays notable endothermal effect, which is coupled to moderate disordering described by an increase of $\alpha$ from 0.42 to 0.52 in the aforementioned temperature range. Disordering of the relaxed sample above $T_g$ is described by the same rise of $\alpha$ as in the case of initial sample. The latter behavior is expected as well since MGs' structural order in the supercooled liquid
state is independent of thermal prehistory (see e.g. Ref.\cite{MakarovMetals2022}). Thus, the evolution of the disorder parameter qualitatively agrees with the character of structure evolution upon heating of initial and relaxed samples.    

\begin{figure}[t]
\center{\includegraphics[scale=0.45]{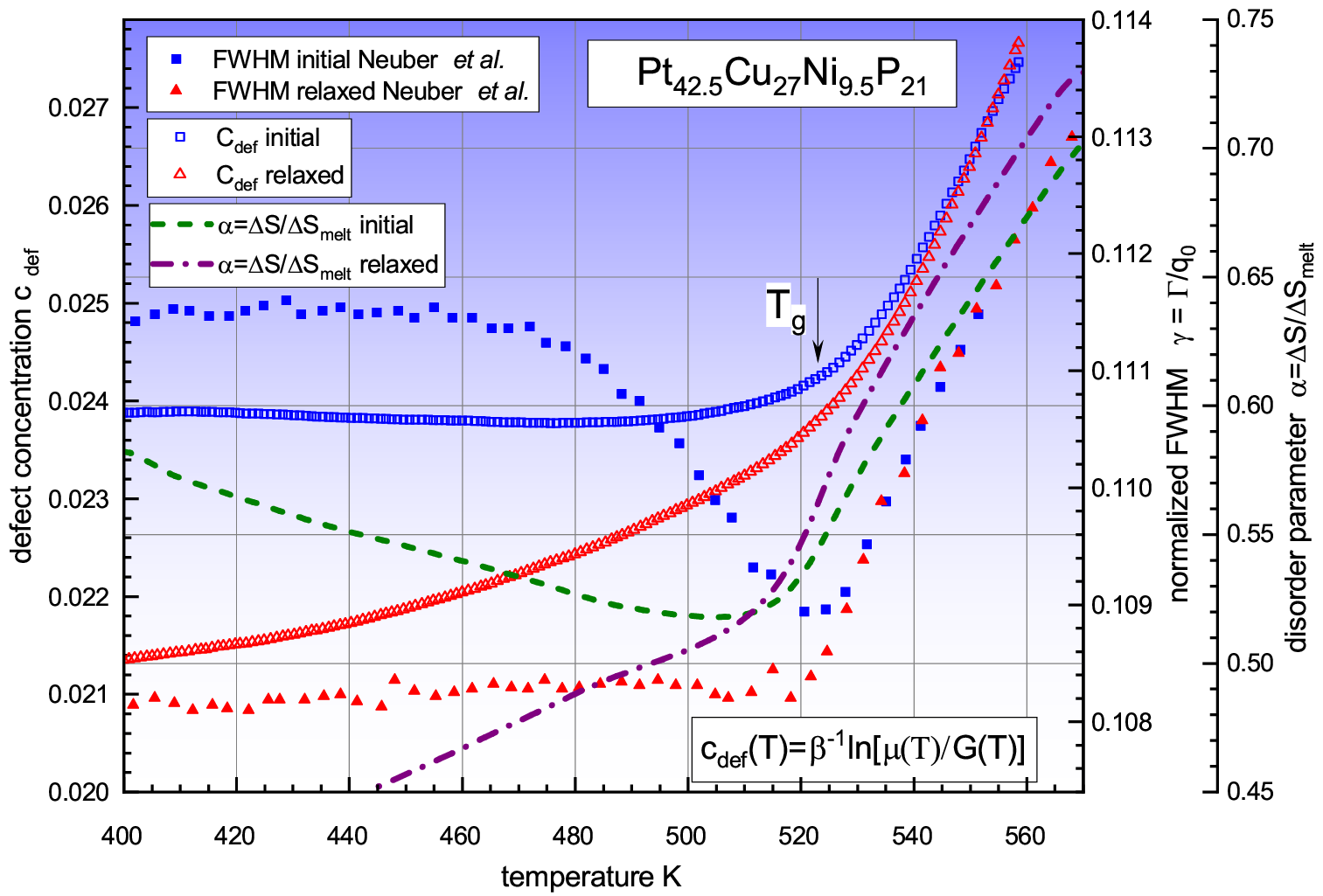}}
\caption[*]{\label{Fig2.eps} Temperature dependences of the FWHM $\gamma$ (original data are taken from the work by Neuber \textit{et al.} \cite{NeuberActaMater2021}) and defect concentration $c_{def}$ reproduced from Ref.\cite{MakarovInmermetallics2023} together with the disorder parameter $\alpha=\Delta S/S_{mel}$ (right axes) shown in Fig.\ref{Fig1.eps}. Calorimetric glass transition temperature $T_g$ (see Fig.\ref{Fig1.eps}) is indicated by the arrow.}
\end{figure}

One can now combine these results with the information on the FWHM and defect concentration data reported earlier in Ref.\cite{MakarovInmermetallics2023}. In this work, the FWHM  was taken as the width $\Gamma$ of the 1st peak of the structure factor $S(q)$ normalized by the scattering vector $q_0$ corresponding to this peak, i.e.  $\gamma=\Gamma/q_0$. The defect concentration was calculated withing the framework of the Interstitialcy theory \cite{KobelevUFN2023} using high-frequency shear modulus data as $c_{def}(T)=\beta ^{-1}ln\left[\mu (T)/G(T)\right]$, where $G(T)$ and  $\mu (T)$ are temperature dependences of the shear modulus of the glass under investigation in the initial state and after complete crystallization, respectively, and dimensionless shear susceptibility $\beta\approx 18$ \cite{MakarovInmermetallics2023}. Temperature dependences of the FWHM $\gamma$, defect concentration $c_{def}$ and disorder parameter $\alpha=\Delta S/\Delta S_{melt}$  given above (see Fig.\ref{Fig1.eps}) are shown in Fig.\ref{Fig2.eps}. 

It is seen that the relationship between $\gamma$, $c_{def}$ and $\alpha$ above $T_g$ is quite clear and straightforward: the disorder parameter $\alpha$, FWHM $\gamma$ and defect concentration $c_{def}$ sharply increase with temperature.  The interpretation of this behavior is  rather obvious. Heating above $T_g$ leads to rapid defect multiplication (increase of $c_{def}$), which is accompanied by strong heat absorption (see the endothermal effect demonstrated by the heat flow $\Delta W$  in Fig.\ref{Fig1.eps} above $T_g$). On the other hand, an increase of the defect concentration rises the amount of regions with damaged short-range order increasing thus the FWHM, as discussed earlier \cite{MakarovInmermetallics2023}. Basing on the results of the present work, one can now argue that an increase of the defect concentration also increases the thermodynamic parameter of disordering $\alpha$ determined by the glass excess entropy.  This is the main finding of the present work. It is also worth emphasizing  that all quantities $\gamma$, $c_{def}$ and $\alpha$ almost do not depend on thermal prehistory, as one would expect. Overall, a quantitative analysis of essentially different phenomena, the shear elasticity, X-ray diffraction and calorimetric behavior of the glass under investigation gives unequivocally consistent results. 

The situation below the glass transition is more complicated but nonetheless quite understandable. A decrease of $\alpha$ (in other words, structural ordering) of initial  samples with temperature well below $T_g$ is conditioned by exothermal structural relaxation (see exothermal $\Delta W$-rise for the initial state in the range 400 K$<T<500$  K in Fig.\ref{Fig1.eps}), which is, however, poorly seen in temperature dependences of the FWHM $\gamma$ and $c_{def}$. Preliminary relaxation of samples by heating into the supercooled liquid state and cooling back to room temperature results in endothermal heat flow even well below $T_g$ (see the same range the range 400 K$<T<500$  K in Fig.\ref{Fig1.eps} for relaxed samples) and related disordering (increase of $\alpha$) even below $T_g$, which is quite well manifested in an increase of the defect concentration  but almost invisible in the FWHM (see $c_{def}(T)$ and $\gamma (T)$ dependences for relaxed samples). Structural disordering with temperature above $T_g$ becomes much stronger. 

In conclusion, we performed calorimetric measurements on bulk glassy Pt$_{42.5}$Cu$_{27}$Ni$_{9.5}$P$_{21}$, which was earlier used to uncover the relationship between the width of X-ray structure factor FWHM $\gamma$ and concentration of defects $c_{def}$ derived using shear modulus data as a function of temperature and thermal prehistory \cite{MakarovInmermetallics2023}. On this basis, we determined the excess entropy of glass with respect to the maternal crystal using Eq.(\ref{DeltaS}) and calculated the thermodynamic parameter of structural disordering $\alpha$ defined by Eq.(1). It was found that the parameter $\alpha$ above the glass transition temperature $T_g$ rapidly increases with the FWHM $\gamma$ and $c_{def}$ proving an interconnection between the thermodynamic excess entropy-based and diffraction measures of disorder and their relation to the defect concentration. Overall, this investigation confirms earlier idea \cite{MakarovInmermetallics2023} that macroscopic structural disordering in metallic glasses above $T_g$ is intrinsically related to their structural defects. Below $T_g$, the interrelation between $\alpha$, $\gamma$ and $c_{def}$ is more complicated but nonetheless follows reasonable expectations.

\section*{Acknowledgments}

The work was supported by the Russian Science Foundation under the project 23-12-00162.

\end{document}